\documentclass[preprint]{aastex}

\slugcomment{Submitted to The Astrophysical Journal}

\shorttitle{ISOCAM observations of globular clusters}
\shortauthors{Origlia et al.}

\begin{document}
 
\title{
ISOCAM\altaffilmark{1} observations of Galactic Globular Clusters: \\
mass loss along the Red Giant Branch.}
  
\author{Livia Origlia and Francesco R. Ferraro}
\affil{Osservatorio Astronomico di Bologna, Via Ranzani 1,
        I--40127 Bologna, Italy}                   
\email{origlia@bo.astro.it, ferraro@bo.astro.it}
 
\author{Flavio Fusi Pecci\altaffilmark{2}}   
\affil{Stazione Astronomica, 09012 Capoterra, Cagliari, Italy} 
\email{flavio@bo.astro.it}
 
\and 

\author{Robert T. Rood}   
\affil{Department of Astronomy, University of Virginia, P.O. Box 3818,
Charlottesville, VA 22903--0818} 
\email{rtr@virginia.edu}

\altaffiltext{1}{Based on observations with ISO, 
     an ESA project with instruments
     funded by ESA Member States (especially the PI countries: France,
     Germany, the Netherlands and the United Kingdom) with the
     participation of ISAS and NASA.}
\altaffiltext{2}{On leave from Osservatorio Astronomico
       di Bologna.}

\begin{abstract}

Deep images in the 10 $\mu$m spectral region have been obtained for five massive
Galactic globular clusters, NGC~104 (=47~Tuc), NGC~362, NGC~5139
(=$\omega$~Cen), NGC~6388, NGC~7078 (=M15) and NGC~6715 (=M54) in the
Sagittarius Dwarf Spheroidal using ISOCAM in 1997.  A significant
sample of bright giants have an ISOCAM counterpart but only $<$ 20\%
of these have a strong mid-IR excess indicative of dusty circumstellar
envelopes.  From a combined physical and statistical analysis we
derive mass loss rates and frequency. We find that {\it
i)}~significant mass loss occurs only at the very end of the Red Giant
Branch evolutionary stage and is episodic, {\it ii)} ~the modulation
timescales must be greater than a few decades and less than a 
million years, and {\it iii)}~mass loss occurrence 
does not show a crucial dependence on the cluster metallicity.
\end{abstract}
 

\section{Introduction}
 
A complete, quantitative understanding of the physics of mass loss processes 
and the precise knowledge of the gas and dust content 
in Galactic globular clusters is crucial in the study of Population 
II stellar systems and their impact on the Galaxy evolution. 
It also has major astrophysical implications on
related problems such as the ultraviolet excess found in elliptical galaxies 
(Greggio \& Renzini 1990; Dorman et al. 1995) and 
the interaction between the intracluster medium and the hot halo gas 
(cf. e.g. Faulkner \& Smith, 1991). 

Indirect evidence for mass loss in Population II stars 
includes the observed morphology of the horizontal branch (HB) in the cluster  
color--magnitude diagrams, the pulsational properties of the RR Lyrae stars, 
and the absence of asymptotic giant branch (AGB) stars brighter than the red 
giant branch (RGB) tip (Rood 1973, Fusi Pecci \& Renzini 1975; 1976, 
Renzini 1977, Fusi Pecci et al. 1993, D'Cruz et al. 1996). 
The expected mass loss is about $0.2\,M_{\odot}$ along the RGB  
and about $0.1\,M_{\odot}$ along the AGB (e.g. Fusi Pecci \& Renzini 1976).
As a consequence of such mass loss processes, dust and gas should be
present in the intracluster medium, as diffuse clouds
(cf. e.g. Angeletti et al. 1982) or concentrated in circumstellar
envelopes.  
If no cleaning mechanism is at work between two Galactic plane crossings, 
a few tens solar masses of intracluster matter should be
accumulated in the central regions of the most massive clusters (i.e. those
with large central escape velocity).                              

Early searches (see Roberts 1988, Smith et al. 1990 and references
therein) for this intracluster gas (HI, HII, CO) 
and more recently detection of ionized gas in 47 Tuc 
(Freire et al. 2001), yielded upper limits or marginal detections 
well below 1~$M_{\odot}$.

Mid--IR excesses and scattered polarized light have also 
been observed in the central region of massive globular clusters
(Frogel \& Elias, 1988; Gillet et al. 1988; Forte \& Mendez 1989;
Minniti et al., 1992; Origlia, Ferraro \& Fusi Pecci 1995, Origlia et
al. 1997b).  They are mainly associated with long period variables
evolving along the AGB.  These preliminary results seem to indicate
that in globular clusters even the brightest IR point sources are
fainter than 1 Jy in the 10 $\mu$m spectral region and most of the emission
comes from warm circumstellar dust with a minor photospheric
contribution. Any diffuse component is much fainter than 10
mJy$\,{\rm arcsec^{-2}}$.

Other evidence of dust in the intracluster medium of globular clusters
comes from the presence of IRAS sources (Lynch \& Rossano 1990, Knapp
\& Gunn 1995, Origlia et al. 1996) in their central regions and from
more recent surveys using ISOPHOT (Hopwood et al. 1999) and ground
based sub-millimeter antennas (cf. e.g. Origlia et al. 1997a, Penny,
Evans \& Odenkirchen 1997, Hopwood et al. 1998).  

However, all these
studies show that globular clusters are deficient in diffuse
intracluster matter and that some mechanism(s) must be at work to
remove the gas and dust which should have accumulated between each
passage through the Galactic plane (cf. e.g. Faulkner \& Smith, 1991).
 
Given this observational scenario the desirability of a deep mid-IR
survey of the central regions of globular clusters in the continuum
and in selected dusty features is obvious.  Such a deep, spatially
resolved photometric survey became possible with the
spectrophotometric capabilities of ISOCAM on board of  
the Infrared Space Observatory (ISO, Kessler et al. 1996). 
ISOCAM (Cesarsky et al. 1996) provided relatively fine spatial resolution,
large field coverage, and high sensitivity in the 10 and 20 $\mu$m
spectral regions.

Ramdani \& Jorissen (2001) performed a deep survey at 12 $\mu$m in the
external regions of 47~Tuc at different distances from the cluster
center, with the specific goal of studying the mass loss during the
AGB evolutionary stage.  They cross-correlated their ISOCAM survey
with the DENIS survey and found dust excess only in two well known bright
variables, confirming the strong link between stellar pulsation
activity and mass loss modulation.

With the goal of studying the mass loss during the RGB evolutionary
stage, a deep survey of the very central regions of six, massive
globular clusters has been made using ISOCAM with two different
filters in the 10 $\mu $m spectral region.  Mid-IR observations have the
advantage of sampling an outflowing gas fairly far from the star
(typically, tens/few hundreds stellar radii).  At 10 $\mu$m the sampled
material typically left the star a few decades previously. Thus mass
loss rates inferred from the mid-IR will smooth the mass loss rate
over a few decades. For astrophysical purposes it is the longterm
average mass loss which is important. Conceptually one could sample
different distances and smoothing times by observing at different
wavelengths, however at this time that is not practical. At longer
wavelengths detectors lack the requisite sensitivity; at shorter
wavelengths the photosphere dominates.

\section{The data set}

Five massive Galactic globular clusters, NGC~104 (=47~Tuc), NGC~362,
NGC~5139 (=$\omega$~Cen), NGC~6388, NGC~7078 (=M15) and NGC~6715 (=M54)
in the Sagittarius Dwarf Spheroidal were observed with ISOCAM between
February and August 1997, (proposals ISO\_GGCS and DUST\_GGC, P.I.:
F. Fusi Pecci).
Each of the clusters has an IRAS point source in its core 
(Origlia, Ferraro \& Fusi Pecci  1996). The 12 $\mu$m IRAS 
flux is 1.7 Jy in 47~Tuc and 
1.2 Jy in NGC~6388 and ranges between 0.4--0.5 Jy in the others.
47~Tuc, NGC~6388, and NGC~5139 show extended IRAS emission at 12 $\mu$m as 
well, with typical sizes of 2--4\arcmin\ and total flux densities between 
3 and 11 Jy. 

Beam-switching observations were performed using CAM03 and
6\arcsec$\,{\rm pixel^{-1}}$ scale in two filters,
LW7[9.6]\footnote{For clarity we give the central wavelength (in
$\mu$m) of the ISOCAM filters in brackets.} (8.3--10.9 $\mu$m,
centered on a silicate dust feature) and LW10[12] (8.6--15.4 $\mu$m,
the IRAS 12 $\mu$m band).  The total integration time for each cluster
was about 2500s and the field of view was about
3\arcmin$\times$3\arcmin.  47~Tuc was also observed a second time in
four filters, LW6[7.7] (6.9--8.5 $\mu$m), LW7[9.6], LW8[11.3]
(10.7-11.9 $\mu$m) and LW10[12], for a total integration time of about
6700s and a field coverage of about 5\arcmin$\times$3\arcmin.

Near-IR photometry of the clusters observed with ISOCAM was obtained
at ESO, La Silla (Chile), using the ESO-MPI 2.2m telescope and the
near IR camera IRAC-2 (Moorwood et al. 1992) equipped with a NICMOS-3
256$\times$256 array detector, during different runs between 1992 and
1994.  The frames were taken through standard $J$ and $K$ broad band
filters.  The overall spatial coverage is about
4\arcmin$\times$4\arcmin\ with a 0\farcs 5 pixel.  More details on the
reduction procedure and the photometric calibration are given by
Ferraro et al. (1994; 2000) and Montegriffo et al. (1995).
$\omega$~Cen was also observed (May 1999) with the ESO-NTT telescope
using SOFI, the near-IR imager/spectrometer equipped with a
1024$\times$1024 array detector. The images were acquired through the
$J$ and $K_s$ filters using a 0\farcs 3 pixel scale. The $K_s$
photometry has been converted into the standard $K$ system using the
stars in common with the IRAC-2 survey, and the calibration of
Origlia, Ferraro \& Fusi Pecci (1995).  Accurate Point Spread Function
(PSF) fitting reductions of the stellar fields have been carried out
using the ROMAFOT package (Buonanno et al. 1983).

Fig.~1 shows the $K$ band images of the six clusters with
the ISOCAM 12 $\mu$m contour plots superimposed. The latter were obtained 
using the background subtracted, mosaiced images (CMOS) generated by 
the ISOCAM auto-analysis pipeline (7.0 version of the off-line 
processing system).

\section{Stellar Point Sources}

The ISOCAM point source catalogs were also generated using the
auto-analysis pipeline.  For each beam--switching pointing the
pipeline generates a calibrated flux image (CMAP) and the
corresponding catalog of detected point sources.  The point source
searching and photometry was performed by maximum-likelihood estimates
and PSF fitting.  The nominal flux uncertainty is $\le 20$\%.

For each filter configuration we cross-correlated the various ISOCAM point
source catalogs from the different beam-switching pointings and
generated a final catalog with average fluxes and positions.
This procedure has also allowed us to remove spurious detections due to
possible transient or memory effects.  The fluxes were first
color--corrected, according to a blackbody energy distribution in the
range of temperatures between 3000 and 5000 K, typical of the globular
cluster red giants.  The applied color--correction factors for the
LW6[7.7], LW7[9.6], LW8[11.3] and LW10[12] filters are 1.06, 1.01, 1.00 and 1.26,
respectively.

In order to convert the observed fluxes into magnitudes, 86.5, 66.4,
43.6 and 28.3 Jy were adopted as zero--magnitude fluxes at the
reference wavelengths of the LW6[7.7], LW7[9.6], LW8[11.3] and
LW10[12] filters, respectively.  These zero point values were obtained
using the IRAS flux density calibration at 12 $\mu$m (cf. Beichman et
al. 1985) and a 10,000~K blackbody to obtain the zero-magnitude fluxes
at the other reference wavelengths.

The near-IR stellar counterpart of each ISOCAM point source was
identified by roto-translating the ISOCAM coordinate system into the
near-IR system and cross-correlating the two catalogs with an 
overall accuracy $\le$2 arcsec. 
Given the
large pixel size of the ISOCAM images, several stars are often
associated to each mid-IR source on the basis of the spatial
coordinates alone. The brightest one within a square box of 
30\arcsec~$\times $~30\arcsec\ (5~$\times $~5 ISOCAM pixels) was
selected.
In all cases the brightest star turns to be also the reddest in (J-K).  
  
The near-IR $J$ and $K$ magnitudes were de-reddened and transformed
into bolometric magnitudes using the reddening and distance moduli
given in Table~1, and the bolometric corrections as a function of the
de-reddened $(J-K)_0$ colors listed by Montegriffo et al. (1998). 
The distance moduli in Table~1 are those from Harris (1996) with a
$+ 0.2$~mag correction, based on the new calibration by Ferraro et
al. (1999).

The average accuracy of the relative photometry is $\le 0.2$ mag in the 
mid-IR and $\le 0.1$ mag in the near-IR.
The uncertainty of the zero point calibrations 
for absolute photometry is somewhat larger, 
a conservative estimate being $\le 0.5$ mag.

In the two closest clusters, 47~Tuc and $\omega$~Cen, the limiting
bolometric magnitude is $M_{\rm bol} \le -1.0$; in M15 and NGC~6388 
(the most distant ones in our Galactic sample) 
the detection threshold is $M_{\rm bol} \le -2.5$.  
Our statistical analysis is based on this second limit. 
M54 in the Sagittarius Dwarf Spheroidal is more distant and even the brightest 
stars are close to the detection limit of 
our ISOCAM survey. 
The number of
stellar counterparts identified in each cluster is given in
Table~2. The $K$ band stellar maps of the six clusters down to a
bolometric magnitude $M_{\rm bol}=-2.5$ and their associated ISOCAM
point sources are plotted in Fig.~2.  The only known variables in this
sample are V42 in $\omega$~Cen (a long period variable) and six other
stars with small amplitude variability and unknown period 
present in the observed field of view of
47~Tuc (cf. Montegriffo et al. 1995 and reference therein).

Using the ISOCAM mosaiced images (CMOS, cf. Fig.~1) we also performed
aperture photometry of the central core region of each cluster within
an $\approx 1\farcm 5$ radius. We then compared these integrated flux
densities with those obtained by summing the contributions of each
ISOCAM point source over the same field of view and determined an upper 
limit to the percentage concentrated in circumstellar (CS) envelopes with
identified stellar counterparts. Both the total fluxes and these
percentages are given in Table~2. 
These percentages scale approximately with the geometrical dilution, 
and very similar values are also inferred using the $K$ band flux, which 
is fully dominated by the stellar photosphere. 
Such a behavior suggests that the 12 $\mu$m emission
not associated with bright circumstellar envelopes is most likely due
to the unresolved stellar background, although some contribution by
diffuse warm dust heated by the cluster radiation field can be also
possible.

\section{Color excess and dust parameters}

Suitable color--magnitude and color--color diagrams of the
ISOCAM point sources were constructed in order to
quantify the relative photospheric and dust excess contribution. 
Fig.~3 shows the $M_{\rm bol},~(J-K)_0$ and $M_{\rm bol}
,~(K-[12])_0$ color--magnitude diagrams. Fig.~4 shows the
corresponding $(K-[12])_0,~(K-[9.6])_0$ color--color diagram.

The $(J-K)_0$ color mainly traces the photospheric
temperature, thus defining the RGB ridge line of each cluster (Ferraro
et al.  2000). Possible circumstellar dust excess is best shown using
a combination of near and mid-IR colors. Overall a good correlation
between the $(K-[12])_0$ and $(K-[9.6])_0$ colors is found, indicating
that both the broad band continuum filter LW10[12] and the narrow band
filter LW7[9.6] centered on a silicate dust feature are good tracers of
dust excess.  However, since the broad LW10[12] filter is somewhat more 
sensitive in detecting mid-IR excess and taking into account the 
actual color scatter along the observed RGB,
$(K-[12])_0\approx 0.65$ was adopted as a conservative, 
border value between pure photospheric emission and significant dust
excess.
Mass loss at a lower rate could be traced by bluer $(K-[12])_0$ colors 
as well, but the observed photometric scatter and the spatial
resolution of our mid-IR survey is not adequate to further constrain 
mass loss in this regime.  

20 out of the 52 ISOCAM sources detected in the upper 1.5 bolometric
magnitudes of the RGB show direct evidence of mid-IR circumstellar
dust excess: 2 variables (V8 and V26, cf. Montegriffo et al. 1995) and
5 normal giants in 47~Tuc, 6 in NGC~6388, 3 in NGC~362, 2 in M15, the
only candidate cluster member in M54 and the long period variable V42 in
$\omega$~Cen (also see Table~2). 
The de-reddened photometry of these sources with dust excess is reported in Table~3.
From pure evolutionary timescale considerations 
(e.g., Renzini \& Fusi Pecci 1988), 
10\% or less of bright giants are expected to be in  the AGB phase.
Since we observed the very central region of each cluster where most of the 
luminous giants should be located, we expect  
less than 5 AGB stars in our global sample, hence when lacking further information 
we neglect them since they have a minor impact on the overall statistics.  
One exception is V42 in $\omega$~Cen:
this star well fits the Period-Luminosity relation of classical Mira variables 
(especially if our slightly larger distance modulus is adopted, cf. 
Feast, Whitelock \& Menzies, 2001) 
and it should be more likely on the AGB evolutionary
stage, so we do not consider it in our further statistical analysis
on the mass loss along the RGB.

In NGC~6388 and in M15, the most distant clusters in our 
Galactic sample, also very centrally concentrated,  
both the total number of ISOCAM sources and the number of those with
mid-IR excess must be regarded as lower limits, since stellar
counterparts to the 12 $\mu$m emission cannot be identified in the
very central core where the crowding is too severe given the ISOCAM
spatial resolution.  

Somewhat surprisingly, no significant mid-IR excess has been found
around any giant star in the observed central region of $\omega$~Cen.
However, this cluster is known to be anomalous in many respects, with
a multi-population RGB (cf. Pancino et al. 2000). More crucially in
the current context, while the cluster is large in absolute size, it
is very low density, and our sample size is small. Taking advantage of
the wide field photometry recently published by Pancino et al. (2000),
it has been possible to estimate that only $<30$\% of the brightest
giants are located in the central region covered by our mid-IR survey,
while in the other clusters we sample almost all of the brightest
giants. In order to obtain a similar statistical significance as in
the other clusters of our survey, a much larger mid-IR spatial mapping
is definitely needed.

In order to estimate the dust parameters a simple model 
using an optically and geometrically thin shell 
at constant temperature and with a $\nu B_{\nu}$ grain emissivity (cf. e.g. Natta \&
Panagia 1976; Origlia, Ferraro \& Fusi Pecci 1996; Origlia et
al. 1997b) has been computed. 
Standard grain radius and density of 0.1 $\mu$m and 3 {\rm g~cm$^{-3}$}, respectively, 
are adopted.

The mid-IR fluxes of those sources showing dust excess have been re-calibrated
according to this modeling, using an average dust temperature of 300 K.
The applied color--correction factors for the
LW6[7.7], LW7[9.6], LW8[11.3] and LW10[12] filters are 1.01, 1.01, 1.00 and 0.95,
respectively.                                               
The fraction of flux due to the photospheric emission has been neglected
since it is small and always within the photometric errors.

The dust temperature has been inferred from the $([9.6]-[12])$ color in 
all clusters, for homogeneity. 
Since 47~Tuc has been also observed in two additional narrow band filters 
(LW6[7.7] and LW8[11.3], cf. Sect.~2), 
the $([7.7]-[11.3])$ and $([7.7]-[9.6])$ colors can be also used to 
derive dust temperatures and to check the consistency with those derived 
from the $([9.6]-[12])$ color.   
Fig.~5 shows that only minor ($<$50~K) systematic differences exist between the 
three methods. 
The inferred values of $T_{\rm dust}$ are between
$\approx 200$ and $\approx 600$ K and the dust equilibrium radii $R_{\rm dust}$ 
between 300 and 40 stellar radii, 
respectively.  The corresponding dust masses
range between a few 10$^{-9}$ and a few 10$^{-7}$ $M_{\odot}$.

Fig.~6 shows that for lower dust temperature
there is a somewhat larger dust mass, while there is not any evident
correlation between the latter and the stellar luminosity.\\
The amount of material available to make dust in RGB stars must be
proportional to the cluster metallicity. Only with helium burning
stages could stars produce their own dust raw material; this should
reach the surface only on the luminous AGB if then.  Indeed, Fig.~6
shows that a large amount of dust has been only detected in the most
metal rich clusters of our sample (47~Tuc and NGC~6388) while the two
metal poor stars of M15 have low dust content.

\section{Circumstellar mass loss}

Empirical mass loss rates can be derived from dust masses by assuming 
a gas-to-dust ratio and a typical timescale for the outflow.

Since more dust should be present in more metal
rich objects one also expects that the gas-to-dust ratio increases when
metallicity decreases (cf. e.g. van Loon 2000).  Assuming a
gas-to-dust mass ratio of 200 in 47~Tuc, the value for the
other clusters is scaled accordingly to their metallicity:

\begin{equation}
\rho = M_{\rm gas}/M_{\rm dust}=200 \times 10^{-({\rm [Fe/H]}+0.76)} 
\end{equation}

An empirical estimate of the epoch of dust ejection can be inferred
from the dust equilibrium radius and assuming an outflow velocity:

\begin{equation}
\tau_{\rm outflow} =R_{\rm dust}/v_{\rm outflow} 
\end{equation}

The outflow velocity should depend on the stellar luminosity and 
the gas-to-dust ratio (cf. e.g. Habing, Tignon \& Tielens 1994).
Larger luminosity leads to a higher momentum flow by the photons,
and thus the kinetic energy and velocity in the flow could
be higher.  For lower gas-to-dust ratio the number of grains per
molecule of gas is larger, and could lead to a higher outflow velocity.  
Adopting the Habing et al. (1994) relation:

\begin{equation}
v_{\rm outflow} \propto L_{\star}^{0.3} ~\rho^{-0.5} 
\end{equation}

The dependence of the outflow velocity on the mass loss rate rapidly 
weakens as the latter increases (Habing et al. 1994), and    
it has been neglected in our computations.

Empirical mass loss rates averaged over the time required by dust to reach 
its equilibrium radius have been inferred using the relation:

\begin{equation}
dM/dt = \rho ~M_{\rm dust}/\tau_{\rm outflow}
\end{equation}

In order to check the impact of our simplified modeling  
on the inferred mass loss rates, several simulations using the DUSTY code
(Ivezi\'c, Nenkova \& Elitzur 1999) have been performed.  A large grid
of values for the overall optical depth at 12 $\mu$m, stellar
temperatures, dust grain sizes, compositions and inner bound
temperatures has been explored, using the analytic approximation to
compute the density distribution for radiatively driven winds.

For homogeneity with our computations an average grain radius of 0.1
$\mu$m has been used, but very similar results (within 10\%) are
obtained with a standard grain size distribution between 0.005 and
0.25 $\mu$m.  An average stellar temperature of 3500 K has been
adopted. Temperature changes of $\pm 500$ K yield overall infrared
excesses, mass loss rates and terminal outflow velocities which are
only $\le 50$\%, $\le 30$\% and $\le 10$\% different, respectively.
Lowering the inner boundary dust temperature from 1000 K (our
reference value) to 500 K, gives a larger (but always within a factor
of 2) variation in the inferred infrared excess, mass loss rates and
outflow velocities. Varying the shell thickness from 10 to 10,000
times the shell inner radius produces negligible variations.

Fig.~7 shows comparisons between the observed infrared excesses, mass
loss rates and outflow velocities inferred by our simplified modeling
and some suitable DUSTY simulations, scaled to the stellar luminosity
and gas to dust ratio of the giants in 47 Tuc.  A good, simultaneous
consistency (within a factor of two) between the empirical and
simulated values have been obtained for an optically thin dust with
$\tau\le 0.05$ and by assuming an outflow velocity of 14 ${\rm
km\,s^{-1}}$ for a stellar luminosity $L_{\star}=1000~L_{\odot}$ and
$M_{\rm gas}/M_{\rm dust}=200$, at the 47~Tuc metallicity.

For the stars with mid-IR excess in our sample
the epochs of dust ejection are typically decades 
and total mass loss rates between 10$^{-7}$ and 10$^{-6}$
$M_{\odot}\,{\rm yr^{-1}}$ (cf. Table~3).  
The systematic uncertainty of these
estimates, mainly due to the various model assumptions, is a factor of
about 3.

Fig.~8 shows that the inferred mass loss rates are almost independent
of the amount of circumstellar dust and stellar luminosity.  Also
there is not any obvious trend with metallicity within the scatter.
Nevertheless, it is worth noticing that while the stars in the
intermediate and metal rich clusters span the whole range of measured
mass loss values, the two metal poor stars of M15 have rather 
high rates, despite the relative low content of dust (cf. Fig.~6).
This is not surprising since one expects that among the most metal poor
objects only those with large mass loss rates can have
detectable dust. The blue tail HB morphology of M15 also shows that
some of its stars (if genuine post-RGB stars) 
have lost almost their entire envelope on the RGB.

These inferred mass loss rates are finally compared with different
empirical laws by Reimers (1975a,b), Mullan (1978), Goldberg (1979),
and Judge \& Stencel (1991) in Fig.~9.  The reference formulae are
taken from Catelan (2000, cf. Appendix and Figure 10), who revised the
original ones taking advantage of the most extensive data set made
available by Judge \& Stencel (1991) and more recent data on
intermediate and low mass giants.  Our rates  are about one order of
magnitude larger than those predicted by these relations and do not 
show any clear dependence on stellar parameters.  However, as
pointed out by Catelan (2000) these relations are calibrated on
Population I giants of relatively low luminosity and extrapolation to
the short-lived evolutionary phases near the tip of the RGB might be
fraught with uncertainty.  These lower mass loss rates could be indeed
consistent with the lower color excess (if any) possibly measured in
47~Tuc and $\omega$~Cen at $M_{bol}\ge-2.5$ (cf. Fig.~3 and Sect.~4).

\section{Discussion}

Dust grains are very efficient in absorbing and scattering starlight
and could contribute to driving or accelerating stellar winds.
However, as widely discussed in the recent review by Willson (2000),
pure stationary flow models seem inadequate to describe the mass loss
phenomena since a strongly dust-driven wind in an oxygen-rich star
requires large light-to-mass ratios ($L/M >10^4\,L_{\odot}/M_{\odot}$). 
Only the coolest and most luminous AGB
stars with direct signatures of dusty winds possibly satisfy this
criterion.  If winds are dust driven, non-steady phenomena like
pulsational effects must be invoked to drive or enhance a stellar wind
and to promote dust formation.  The role of pulsating atmospheres in
AGB mass loss has been considered for many years both on theoretical
and observational basis (cf. Willson 2000 and references
therein). 

For stars on the RGB basic mass loss mechanisms are still quite
unknown (cf. e.g. Judge \& Stencel 1991) and matter of debate.  There
is observational evidence of small-amplitude and short period
variability among giants close to the RGB-tip (cf. e.g. Welty 1985;
Edmonds \& Gilliland 1996), consistent with some oscillation activity
(low-overtone radial or nonradial pulsations).  Cacciari \& Freeman
(1983) and Gratton, Pilachowski \& Sneden (1984), observed H$\alpha$
emission lines in hundreds of giants in a large sample of globular
clusters. They initially argued that the H$\alpha$ emission was direct
evidence for mass loss. However, interpreting H$\alpha$ emission can
be complicated (Dupree 1986). For example, Dupree et al. (1994)
suggest from its time variability that H$\alpha$ emission is due to
atmospheric shocks which might drive the mass loss. In either
interpretation the presence of H$\alpha$ emission in
RGB spectra would generally indicate ongoing mass loss.

In the star in which mass loss is best understood, the Sun, the mass
loss mechanism is ultimately tied to the magnetic field generated by
the convective zone (e.g., Dupree 1986). Since cluster RGB stars have
convective zones they could well have magnetic fields. There are no
direct measurements of such fields, and interpreting signatures of
chromospheric activity which might arise from magnetic fields is quite
complex (e.g., Ayres et al. 1997). There is some recent evidence for
RGB starspots which would imply magnetic activity (Stefanik et
al. 2002). Soker (2000) and Garc\'\i a-Segura et al. (2001) argue that
magnetic activity might play a role in AGB mass loss on the basis of
planetary nebula morphology. 

Some non-canonical
deep mixing in the upper RGB has been also invoked to produce excess
luminosities and enhancing the mass loss close to the RGB-tip
(cf. e.g. Sweigart 1997; Cavallo \& Nagar 2000, Weiss, Denissenkov \&
Charbonnel 2000), but a solid physical picture of the overall impact
on the red giant evolution is still lacking.

Can our observations give further information on the RGB mass loss
process?

According to our mid-IR survey, strong mass loss processes seem to
occur only close to the RGB-tip, during the last few million years of
the RGB lifetime at $M_{\rm bol}\le -2.5$ (Fusi Pecci \& Renzini 1975;
Salaris \& Cassisi 1997).  Moreover, only a fraction of giants
detected by ISOCAM show strong mid-IR excess.  This is especially
clear in Fig.~3. Except near the very RGB-tip, stars with mass loss
are scattered pretty uniformly along the RGB along with stars with no
mass loss. Further, there is no smooth variation with the stellar
luminosity. This suggests episodic mass loss processes.  We will refer
to the fraction of the time that mass loss is ``on'' as $f_{\rm
ON}$. The fraction of time that the mass loss is ``off,'' or has
dropped below our detection threshold, is $f_{\rm OFF}$. Since of the
52 ISOCAM sources 20 show evidence for mass loss, a first global
estimate is $f_{\rm ON} \approx 0.4$. Considering the small numbers,
there is no evidence that $f_{\rm ON}$ for the individual clusters
varies from this global value.

Things are not so simple.  In principle, all the stars brighter than
$M_{\rm bol}\le -2.5$ in each cluster should have been detected.  In
practice, due to the broad ISOCAM PSF, some stars are missed. First,
while there is sometimes more than one star within an ISOCAM point
source PSF, we selected only the brightest. Second, in the dense cores
of M15 \& NGC~6388 it is not possible to associate the ISOCAM objects
with any near-IR stellar counterpart. The number of lost stars
basically depends on the crowding, and thus on both the cluster
distance and density---the larger their values the larger the number
of stars in each pixel field of view, other things being constant.
 
To estimate the number of missed stars, we have computed the expected
number of giants brighter than $M_{\rm bol}\le -2.5 $ within the
ISOCAM field of view in two ways. First we used a theoretical
approach: the fraction of light within the
ISOCAM beam, was computed using a King model and the {\it Specific
Evolutionary Flux} by Renzini \& Buzzoni (1986), assuming an average
lifetime $\le 6$ Myr at $M_{\rm bol}\le -2.5$ (Fusi Pecci \& Renzini
1975; Salaris \& Cassisi 1997). As an alternative we counted the number of
stars detected in our $J,~K$ survey within the same beam.
The two methods provide similar results (within 20\%). The numbers
are given in Table~2. These values are significantly larger than the
corresponding number of ISOCAM point sources detected in each cluster,
as expected given the wide ISOCAM PSF.

How can the missed stars affect our estimate of $f_{\rm ON}$? The
missed stars may or may not be losing mass. Only if they differ in
some systematic way from the detected stars will they affect $f_{\rm
ON}$. Since the brightest and reddest star was selected and since
strong mass loss occurs only close to the tip of the RGB, the missed
(fainter) stars may be less likely to have mass loss. A lower limit
can be obtained by assuming that none of the missed stars has
detectable mass loss. In that case for each cluster $f_{\rm ON} \ge$
(number of stars with dust excess)/(total number of stars) (cf. last
line of Table~2).

In 47~Tuc and NGC~362, where crowding does not severely affect statistics,  
$f_{\rm ON}\ge$15\%, while lower values have been inferred in  M15 and NGC~6388  
but in this cluster the very central core region was not sampled.
In $\omega$~Cen 3--4 stars with dust excess should have been detected  
if $f_{\rm ON}\ge$15\% as in 47~Tuc and NGC~362, but taking into account 
that only $<30$\% of the brightest giants are sampled (see Sect.~3), 
the expected number of detections drops down to $\le1$, fully consistent with 
the inferred zero value.
The result of M54 is not statistical significant, since it is too distant 
(see also Sect.~3).

Broad time limits can be placed on the episodic variations.  
At average rates between 10$^{-7}$ and 10$^{-6}$ $M_{\odot}\,{\rm yr^{-1}}$ 
(cf. Table~3) each major mass loss event should
occur on time scales somewhat shorter than the evolutionary times (5--6
Myr on the upper RGB) and longer than the time it takes the outflowing
gas to reach the radii where we observe the dust, i.e., decades, 
whatever be the actual mass loss mechanism (several bursting episodes with
multiple shells or more continuous flows). 
In principle such a lower limit to the episodic variation timescale 
can be as short as an instantaneous event, possibly at much higher 
mass loss rates, but our observations do not allow to constrain 
timescales with such a temporal resolution. 

The total mass lost is crudely  
\begin{equation}
\Delta M \sim {\rm rate} \times \Delta t_{\rm ON} 
\end{equation}
where $\Delta t_{\rm ON} $ is the total time interval 
in which a star experiences mass loss
(either through many short-living episodes or a more continuous
flowing).

In order to account for a total
mass loss of about $0.2\,M_\odot$ required by the models to explain
the HB morphology, and  
by assuming an average ${\rm rate} \approx 3 \times 10^{-7} M_\odot\,{\rm yr^{-1}}$, 
$\Delta t$ should be $\approx 7 \times 10^{5}\,{\rm yr}$, well within 
the inferred limits and almost one order 
of magnitude smaller than the total evolutionary time on the upper RGB. 

Another interesting result seen from Figs.~8,~9 is that there is not any
clear dependence of mass loss on metallicity, although more data, in
particular for the metal poor clusters, are needed to draw firm
conclusions.  Again this result is not unexpected.  
The earlier
H$\alpha$ studies also suggested little metallicity dependence in mass
loss rates and  are consistent with $f_{\rm ON} \sim
0.5$.  However our observed time scales are much longer than the day
to day variation observed in H$\alpha$ emission.  Both the time scale
and $f_{\rm ON}$ could be consistent with RGB magnetic cycles.
Further, studies of horizontal branch morphology (Catelan
\& de Freitas Pacheco 1993; Lee, Demarque \& Zinn 1994) suggest at
most a mild metallicity dependence on mass loss efficiency except
possibly at the lowest metallicities.  In a recent paper van Loon
(2000) analyzing the mass loss rates of a sample of obscured AGB stars
also finds very modest dependence on metallicity, but possibly for 
different reasons related to the pulsation activity (e.g. van Loon 2001). 

The observed metallicity independence of mass loss coupled with the
fact that mass loss for RGB stars can be far below the criterion
for radiatively driven winds suggests that dust may not be
related to the driving mechanism for RGB mass loss.
Dust can simply be a tracer of the out flowing gas and
not responsible for the outflow.

\section{Conclusions}

Our ISOCAM survey of the central region of six massive globular
clusters provided mid-IR photometry for 78 sources with identified
stellar counterparts.  52 sources are associated with bright giants
close to the RGB-tip. Of these about 40\% have strong mid-IR excess
ascribed to the presence of dusty circumstellar envelopes. Correcting
for stars not detected because of the low spatial resolution of
ISOCAM, dusty envelopes are inferred around about 15\% of the brightest 
giants ($M_{\rm bol}\le-2.5$). 

The inferred mass loss rates are in the range $10^{-7} < dM/dt <
10^{-6} M_{\odot}\,{\rm yr^{-1}}$, assuming as reference values an
outflow velocity of 14~${\rm km\,s^{-1}}$ and a gas-to-dust mass ratio
of 200 at the metallicity of 47~Tuc and for a stellar luminosity of 
1000~$L_{\odot}$.

The major astrophysical implications are: 
\begin{itemize}
\item The mass loss occurs very near the RGB-tip and is episodic.

\item
The mass loss episodes must last longer than a few decades and less
than a million years.

\item
There is no indication for a strict metallicity dependence of the
frequency of mass loss occurrence.
\end{itemize}

\acknowledgments

The research has been supported by the Agenzia Spaziale Italiana
through the grants 9920015AS and I/R/27/00.  RTR is partially
supported by NASA LTSA grant NAG 5-6403 and STScI grant GO-8709.  We
thank Jacco van Loon, the Referee of our paper, for the many
suggestions and comments which improved the overall presentation and
analysis of our results.  We thank M. Bellazzini, G. Bono,
C. Cacciari, M. Catelan, Andea Dupree, Jeff Linsky, and P. Persi for
helpful comments and discussions.

\clearpage

%\section{Figure Captions}

\begin{deluxetable}{lcccccccc}
\footnotesize
\tablewidth{17truecm}
\tablecaption{Adopted parameters for the observed globular clusters.} 
\tablehead{
\colhead{Cluster} &
\colhead{$\alpha$ (2000)} &
\colhead{DEC (2000)} &
\colhead{X$^a$} &
\colhead{Y$^a$} &
\colhead{[Fe/H]} &
\colhead{$d$ [kpc]} &
\colhead{E(B-V)} &
\colhead{(m-M)$_V$} 
}
\startdata
NGC104  (= 47Tuc)       & 00~24~05 &--72~04~51 & 30 &218 &  $-0.76$ &  4.5 & 0.04 & 13.41 \\
NGC362                  & 01~03~14 &--70~50~54 &229 &240 & $-1.16$  &  8.7 & 0.05 & 14.86 \\
NGC5139 (= $\omega$Cen) & 13~26~46 &--47~28~37 &453 &508 & $-1.57$  &  5.4 & 0.12 & 14.04 \\
NGC6388                 & 17~36~17 &--44~44~06 & -6 &223 & $-0.60$  & 12.0 & 0.38 & 16.57 \\
NGC6715 (= M54)         & 18~55~03 &--30~28~42 &254 &262 & $-1.59$  & 27.9 & 0.15 & 17.69 \\
NGC7078 (= M15)         & 21~29~58 & +12~46~19 &138 & 93 & $-2.22$  & 10.9 & 0.09 & 15.47 \\
\enddata
\tablecomments{\\
$^a$~Coordinates in pixel units (cf. Sect.~2 and Fig.~2) of the cluster center.}
\end{deluxetable}

\begin{deluxetable}{lcccccc}
\footnotesize
\tablewidth{15.7truecm}
%\tablewidth{\textwidth}
\tablecaption{ISOCAM point sources in the observed globular clusters.} 
\tablehead{
\parbox[t]{2in}{Cluster (NGC)} &
104  &362  &5139 &6388 &7078 &6715 \\
\parbox[t]{2in}{} &
 47Tuc        &              & $\omega$Cen &              & M15         & M54  
}
\startdata
\parbox[t]{3in}{Stellar Counterparts at $M_{\rm bol} \le -2.5$} &
 27 &  5 & 6 &  9 &  4 &  1 \\
\parbox[t]{3in}{Stellar Counterparts at $M_{\rm bol} \le -1.0$} &
 40    & \dots &  19    & \dots & \dots & \dots \\
\parbox[t]{3in}{Integrated central flux (Jy)at 12 $\mu$m } &
 4.5  & 0.6  & 1.43 & 1.70 & 0.5  & 0.6  \\
\parbox[t]{3in}{\% of central 12 $\mu$m flux in resolved stars} &
 45 & 22 & 26 & 8 & 12 & 2  \\
\parbox[t]{3in}{Non-variable stars with dust excess} &
 5 & 3 & 0 & 6 & 2 & 1 \\
\parbox[t]{3in}{Variable stars with dust excess} &
 V8, V26 & \dots & V42   & \dots & \dots & \dots \\
\parbox[t]{3in}{Estimated total of stars at $M_{\rm bol}\leq-2.5$} &
 40 & 20 & 20 & 90 & 40 & 70 \\
\parbox[t]{3in}{\% of stars with dust excess at $M_{\rm bol}\leq-2.5$} &
 $\ge $18 & $\ge $15 & $>$0 & $> 7$ & $>5$ & $>1$ \\
\enddata
%\tablecomments{}
\end{deluxetable}

\begin{deluxetable}{llcccccccc}
\footnotesize
\tablewidth{17truecm}
\tablecaption{Infrared photometry and mass loss rates for the ISOCAM point sources with dust excess.}
\tablehead{
\colhead{Cluster} &
\colhead{star$^a$} &
\colhead{X$^b$} &
\colhead{Y$^b$} &
\colhead{$K_0$} &
\colhead{$(J-K)_0$}&
\colhead{$(K-12)_0$}&
\colhead{$M_{bol}$}&
\colhead{$T^d_{dust}$ [K]}&
\colhead{$\dot M^c$ [$M_{\odot}~yr^{-1}$]} 
}
\startdata
47~Tuc       & \#1 &     55    & 186  &  6.74  &  0.96  &  1.20  &  $-3.92$   & 402  & $2.5\times 10^{-7}$\\
             & \#2 &      8    & 181  &  7.78  &  0.99  &  0.70  &  $-2.84$  &  343  & $8.5\times 10^{-8}$\\
             & \#3 &    $-7$   & 223  &  7.34  &  1.00  &  0.76  &  $-3.26$  &  491  & $1.0\times 10^{-7}$\\
             & \#4 &     99    & 141  &  7.74  &  1.00  &  1.39  &  $-2.86$  &  291  & $2.1\times 10^{-7}$\\
             & \#5 &    $-131$ & 276  &  7.15  &  1.12  &  2.14  &  $-3.28$  &  234  & $1.1\times 10^{-6}$\\
             &  V8 &      3    & 330  &  6.83  &  1.23  &  1.64  &  $-3.46$  &  312  & $4.8\times 10^{-7}$\\
             & V26 &      9    & 257  &  6.55  &  1.24  &  1.16  &  $-3.73$  &  343  & $3.4\times 10^{-7}$\\
 & & & & & & & & & \\
NGC~362      & \#1 &    277    & 248  &  8.94  &  0.96  &  1.75  & $-3.12$  &   335  & $4.4\times 10^{-7}$\\
             & \#2 &    232    & 232  &  8.73  &  0.99  &  1.11  & $-3.29$  &   339  & $2.8\times 10^{-7}$\\
             & \#3 &    224    & 156  &  8.80  &  1.25  &  0.99  & $-2.87$  &   271  & $3.6\times 10^{-7}$\\
 & & & & & & & & & \\
$\omega$~Cen & V42 &    438    & 686  &  8.51  &  0.89  &  2.35  & $-2.58$  &   347  & $7.1\times 10^{-7}$\\
 & & & & & & & & & \\
NGC~6388     & \#1 &     17    & 308  &  9.62  &  0.85  &  0.98  & $-3.30$  &   291  & $1.4\times 10^{-7}$\\
             & \#2 &   $-89$   & 345  &  9.34  &  0.98  &  0.76  & $-3.39$  &   297  & $1.4\times 10^{-7}$\\
             & \#3 &   $-47$   & 291  &  8.86  &  1.08  &  0.75  & $-3.73$  &   281  & $2.2\times 10^{-7}$\\
             & \#4 &     89    & 274  &  9.74  &  1.08  &  1.04  & $-2.85$  &   335  & $1.2\times 10^{-7}$\\
             & \#5 &     34    & 266  &  9.11  &  1.16  &  1.36  & $-3.37$  &   209  & $6.8\times 10^{-7}$\\
             & \#6 &   $-24$   & 164  &  9.06  &  1.25  &  0.90  & $-3.31$  &   276  & $2.3\times 10^{-7}$\\
 & & & & & & & & & \\
M15          & \#1 &    105    & 177  & 10.47  &  0.63  &  1.97  & $-2.67$  &   397  & $3.3\times 10^{-7}$\\
             & \#2 &    150    &  60  &  9.89  &  0.66  &  1.14  & $-3.18$  &   572  & $4.3\times 10^{-7}$\\
 & & & & & & & & & \\
M54         & \#1 &     222    & 156  & 11.17  &  0.71  &  2.30  & $-3.81$  &   272  & $1.1\times 10^{-6}$\\ 
\enddata
\tablecomments{\\
$^a$~Arbitrary numbering.\\
$^b$~Coordinates in pixel units 
(the scale is 0.5\arcsec$\,{\rm pixel^{-1}}$ for all clusters but $\omega $~Cen 
which has 0.3\arcsec$\,{\rm pixel^{-1}}$, cf. Sect.~2).\\
$^c$~Dust temperature assuming a $\nu B_{\nu}$ dust emissivity model (cf. Sect.~4).\\
$^d$~Mass loss rates, by assuming reference outflow velocity of 14 
${\rm km\,s^{-1}}$ and gas-to-dust mass ratio of 200 at the metallicity of 47~Tuc (cf. Sect.~4) 
and for a stellar luminosity of~1000 $L_{\odot}$.
}
\end{deluxetable}

\begin{figure}
\epsscale{1.2}
\plotone{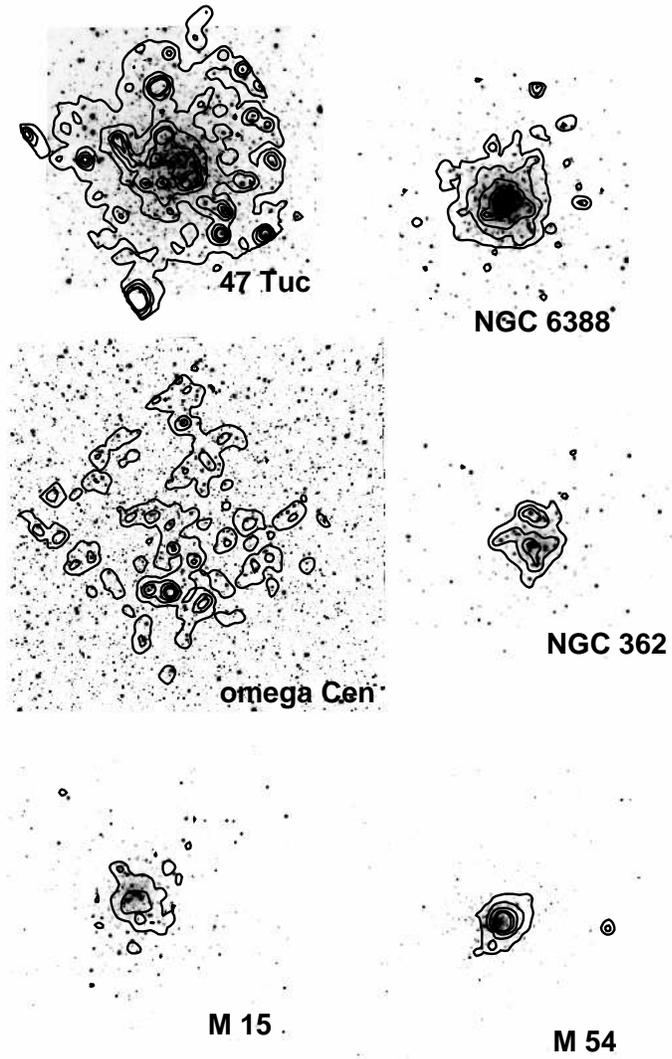}       
\caption{$K$ band images of the six clusters with the 
ISOCAM 12 $\mu$m emission contours at 2, 5, 10, 20 and 30 mJy 
superimposed. North is up, east is left.
Each image field of view centered at the cluster center is 
about 4\arcmin$\times$4\arcmin.
The bright, isolated 12 $\mu$m source on right side from the M54 center is 
associated with a Sagittarius field variable.}
\end{figure}  

\begin{figure}
\epsscale{1.0}
\plotone{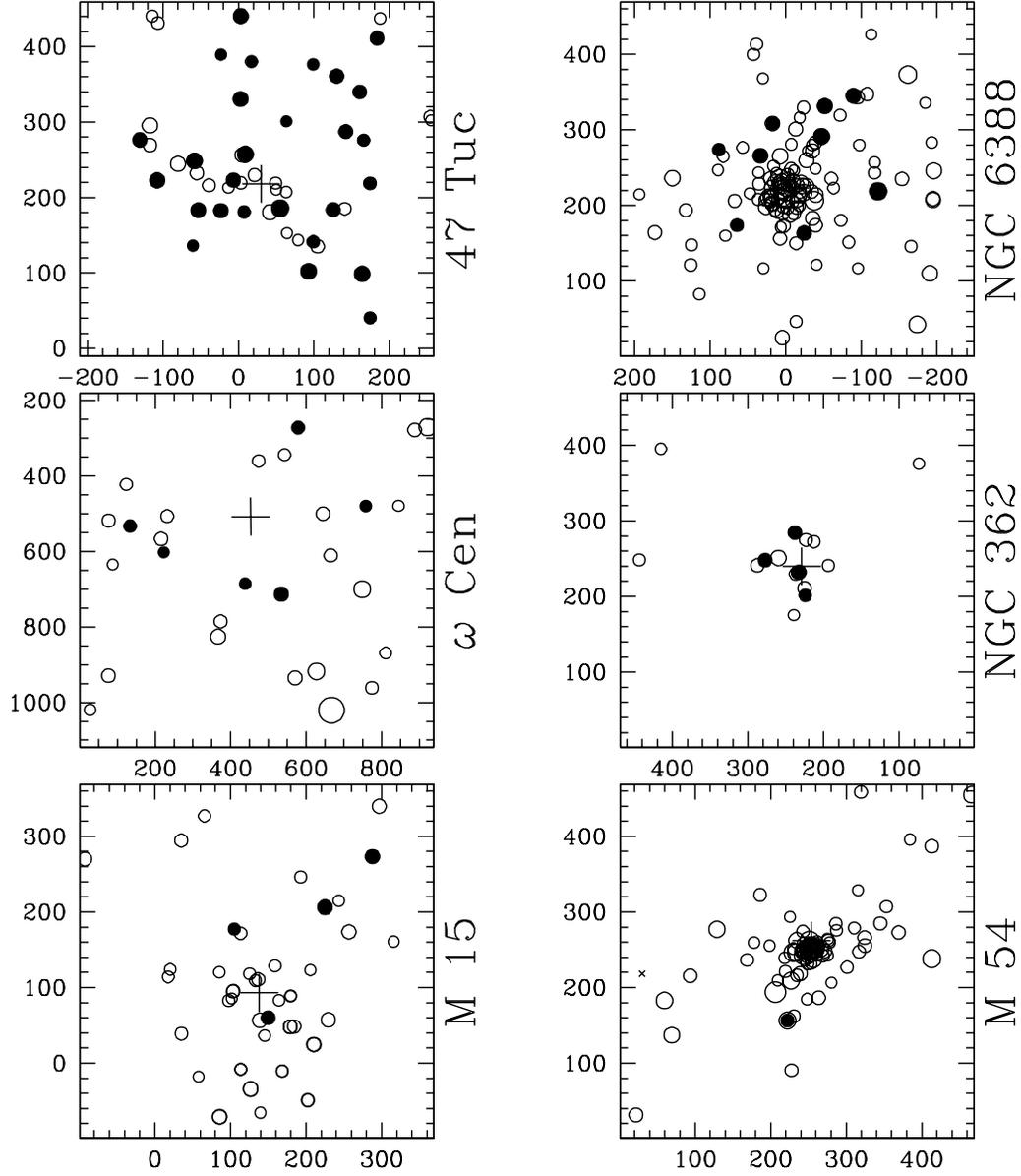}       
\caption{$K$ band stellar maps of the six clusters down to 
a bolometric magnitude $M_{\rm bol}\le -2.5$.
The ISOCAM 12 $\mu$m point source counterparts are overplotted (solid circles). 
North is up, east is left. The X and Y coordinates are in pixel units 
(the scale is 0.5\arcsec$\,{\rm pixel^{-1}}$ for all clusters but $\omega $~Cen 
which has 0.3\arcsec$\,{\rm pixel^{-1}}$, cf. Sect.~2).
Each map field of view centered around the cluster center 
is about 4\arcmin$\times$4\arcmin.
The cross indicates the nominal cluster center (cf. Table~1).}
\end{figure}  

\begin{figure}
\epsscale{1.0}
\plotone{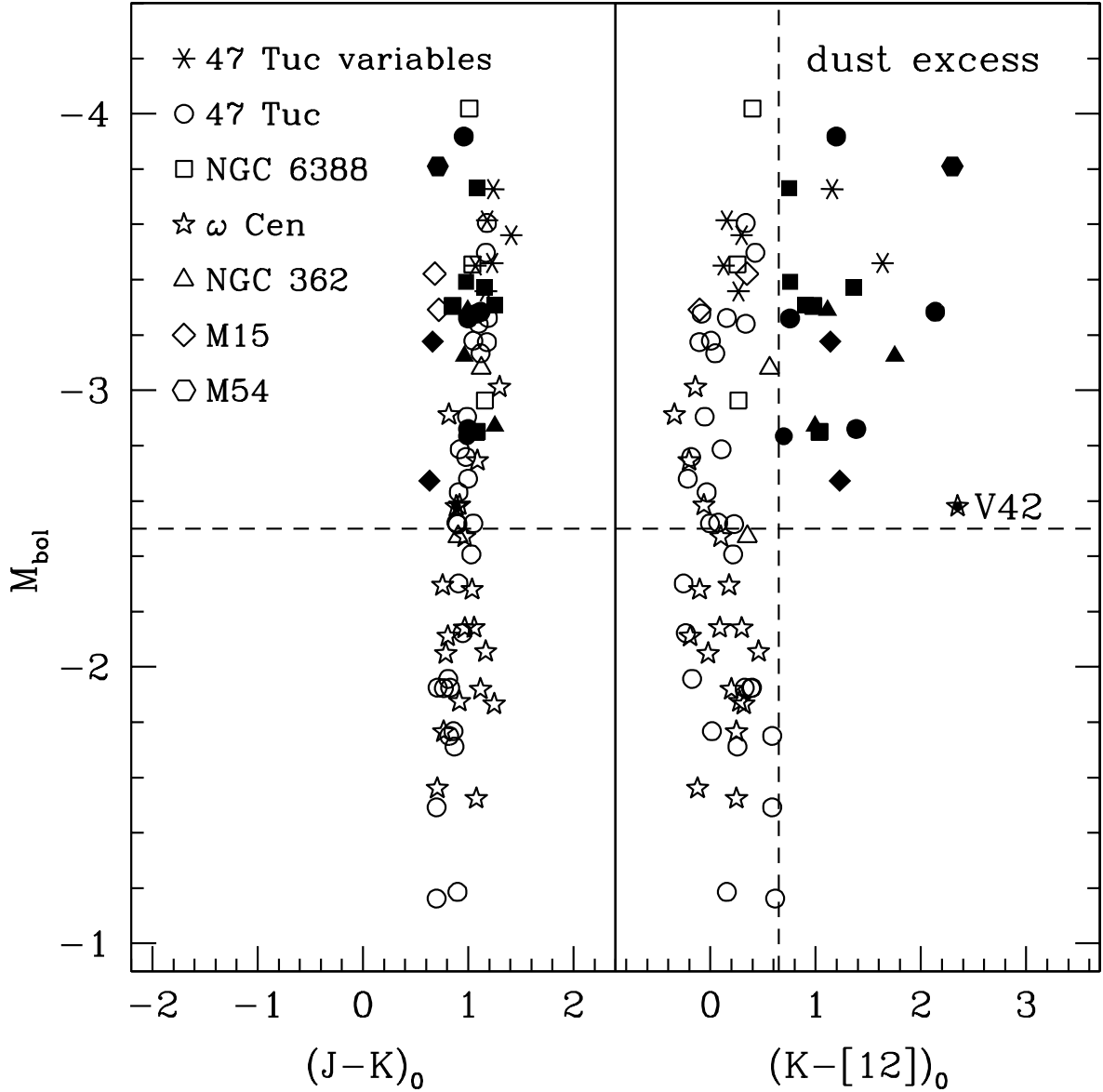}       
\caption{$M_{\rm bol}$,$~(J-K)_0$ (left panel) and $M_{\rm bol}$,$~(K-[12])_0$ 
(right panel) de-reddened color--magnitude diagrams down to 
a bolometric magnitude $M_{\rm bol}$ $\le -1.0$ of the ISOCAM point sources 
detected in the observed globular clusters (cf. Sect.~4).
Sources with $(K-[12])_0\ge 0.65$ are classified as sources with significant 
dust excess and are marked with filled symbols.  
The horizontal, dashed line at $M_{\rm bol} = -2.5$ marks the photometric threshold
in the most distant clusters.
The vertical, dashed line in the right panel marks the border
between where 12 $\mu$m emission is dominated by the stellar photosphere or by 
circumstellar dust.
The position of the V42 long period variable of $\omega$~Cen is also 
marked.}
\end{figure}  

\begin{figure}
\epsscale{1.0}
\plotone{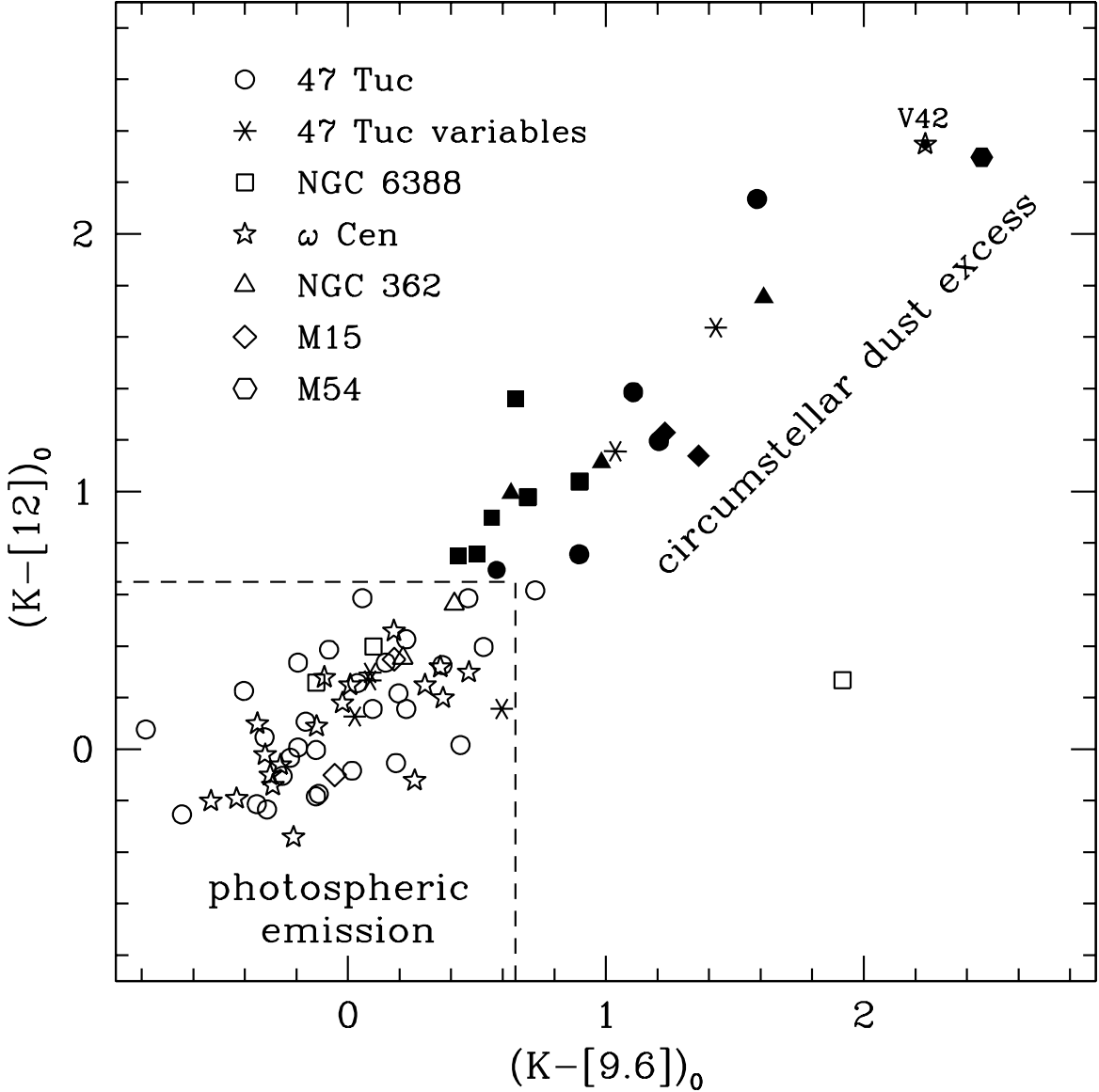}       
\caption{De-reddened $(K-[12])_0,~(K-[9.6])_0$ color--color diagram of the ISOCAM
point sources detected in the observed globular clusters
(cf. Sect.~4). Symbols as in Fig.~3.  The dashed box indicates the
region where the 12 $\mu$m emission is still dominated by the stellar
photosphere.  Although there is a good overall correlation between the
$(K-[12])_0$ and $(K-[9.6])_0$ colors, as a conservative
approach only those sources with $(K-[12])_0>0.65$ are classified as
having dusty envelopes (we then exclude the star in NGC~6388 with 
strong $(K-[9.6])_0$ color excess and photospheric $(K-[12])_0$,
cf. Sect.~4).  
The position of the V42 long period
variable in $\omega$~Cen is also marked.}
\end{figure}  

\begin{figure}
\epsscale{1.0}
\plotone{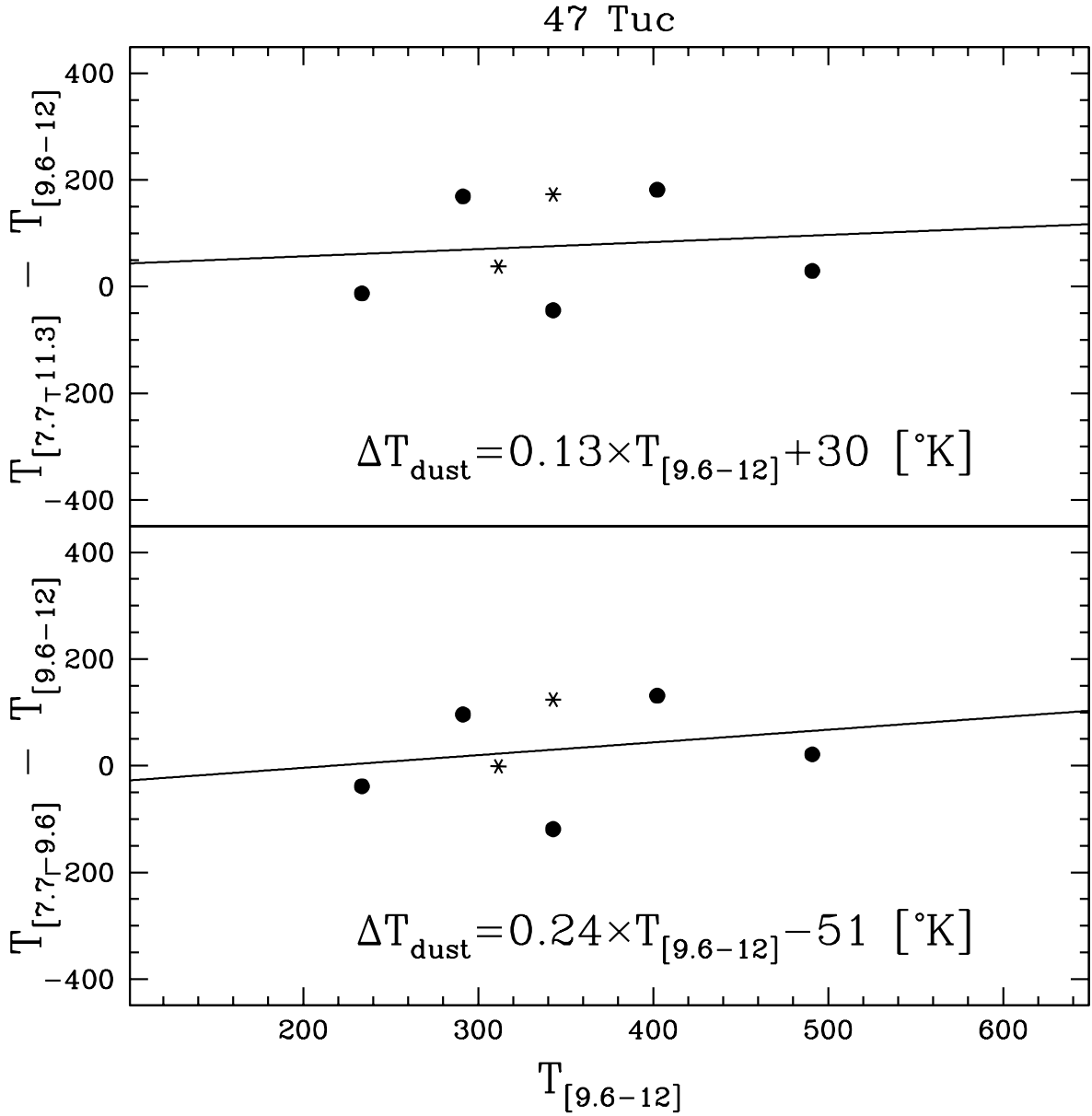}       
\caption{Dust temperatures from different 
mid--IR colors $([7.7]-[12])$, $([7.7]-[11.3])$ and $([7.7]-[9.6])$ 
of the ISOCAM point sources with significant circumstellar dust excess
(cf. Sect.~4) detected in 47~Tuc, assuming a $\nu B_{\nu}$ dust emissivity 
model.  Symbols as in Figs.~3,4.}
\end{figure}  

\begin{figure}
\epsscale{1.0}
\plotone{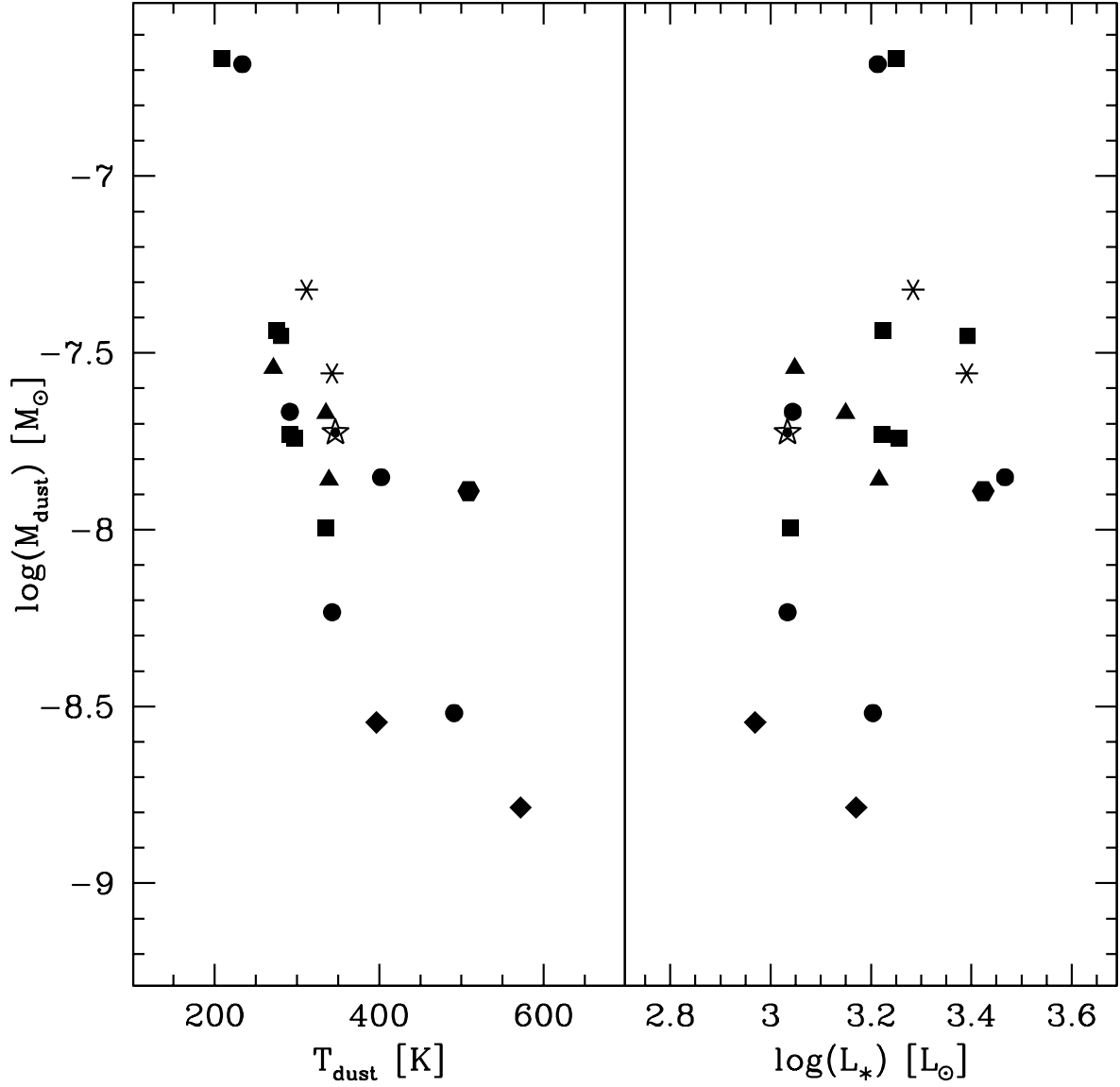}       
\caption{
Dust mass as a function of dust temperature (left panel) and  
stellar luminosity (right panel) for the giant stars with dust excess 
(cf. Sect.~4). 
The position of the V42 long period variable of $\omega$~Cen is also 
marked. Symbols as in Figs.~3,4.
}
\end{figure}  

\begin{figure}
\epsscale{1.0}
\plotone{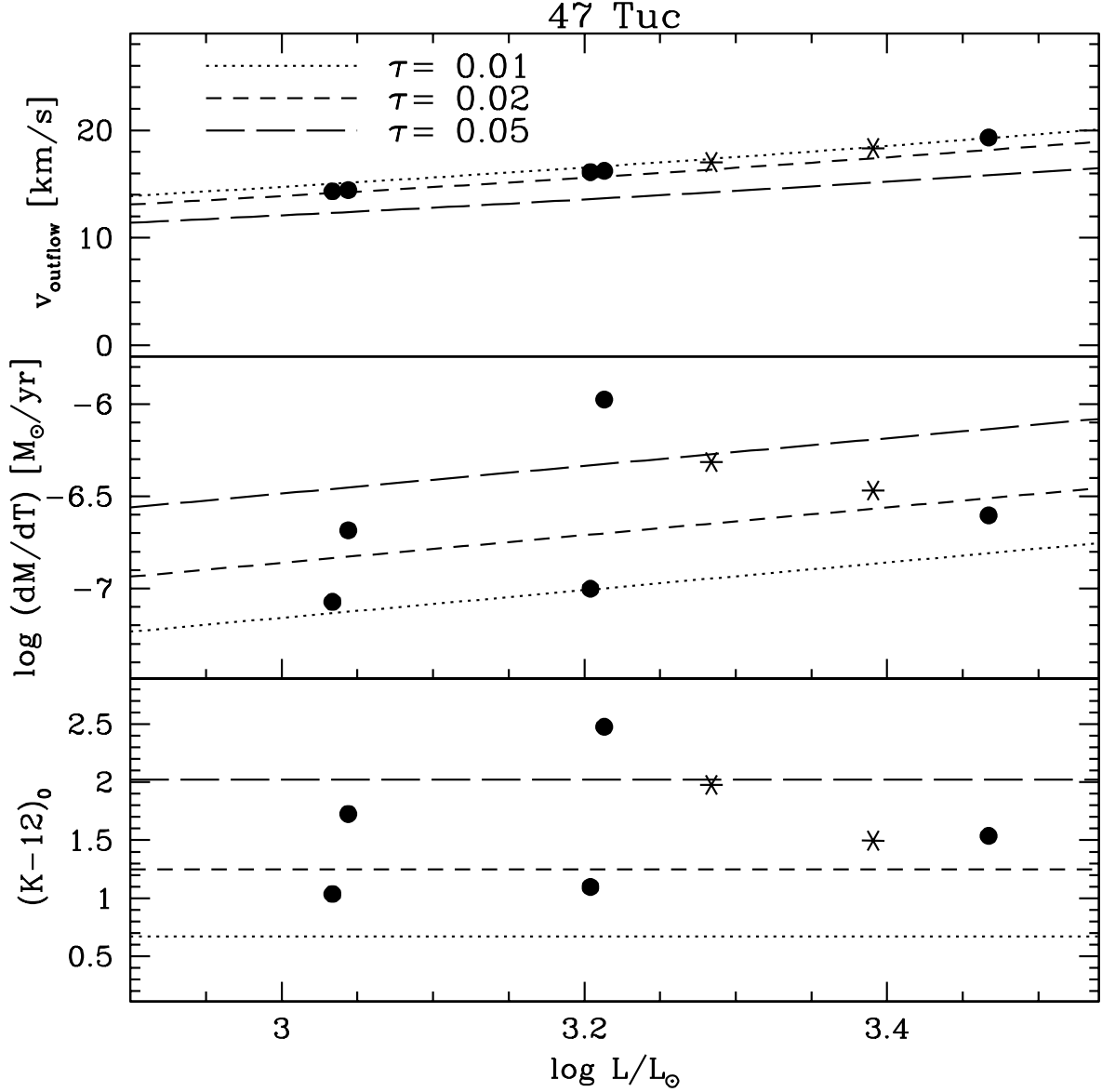}       
\caption{
$(K-12)_0$ infrared excesses (lower panel), mass loss rates (middle panel) 
and outflow velocities (top panel) 
for the 7 giants in 47 Tuc with dust excess. Symbols as in Figs.~3,4. 
The dashed lines refer to DUSTY simulations at different optical depths, 
using as reference parameters an average stellar temperature of 3500 K,
a grain radius of 0.1 $\mu$m, an inner boundary dust temperature of 1000 K and  
a shell outer radius 1000 times the shell inner radius (cf. Sect.~5).
}
\end{figure}  

\begin{figure}
\epsscale{1.0}
\plotone{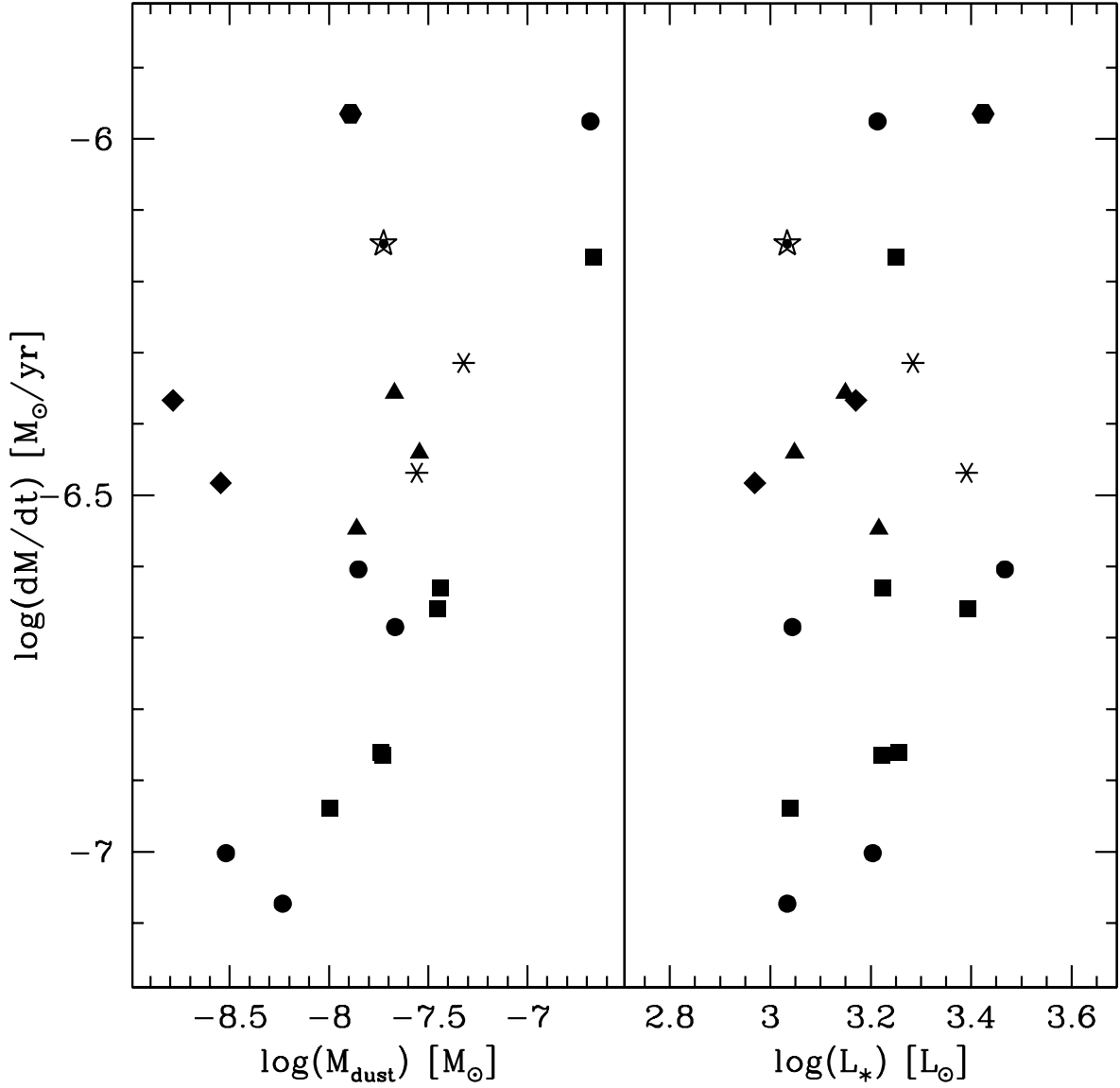}       
\caption{
Total mass loss rates as a function of dust mass (left panel) and  
stellar luminosity (right panel) for the giant stars with 
dust excess (cf. Sect.~4), by assuming  
reference outflow velocity of 14~${\rm km\,s^{-1}}$ and gas-to-dust mass ratio of 200 
at the metallicity of 47~Tuc and for a stellar luminosity of~1000 $L_{\odot}$. 
The position of the V42 long period variable of $\omega$~Cen is also 
marked. Symbols as in Figs.~3,4.
}
\end{figure}  

\begin{figure}
\epsscale{1.0}
\plotone{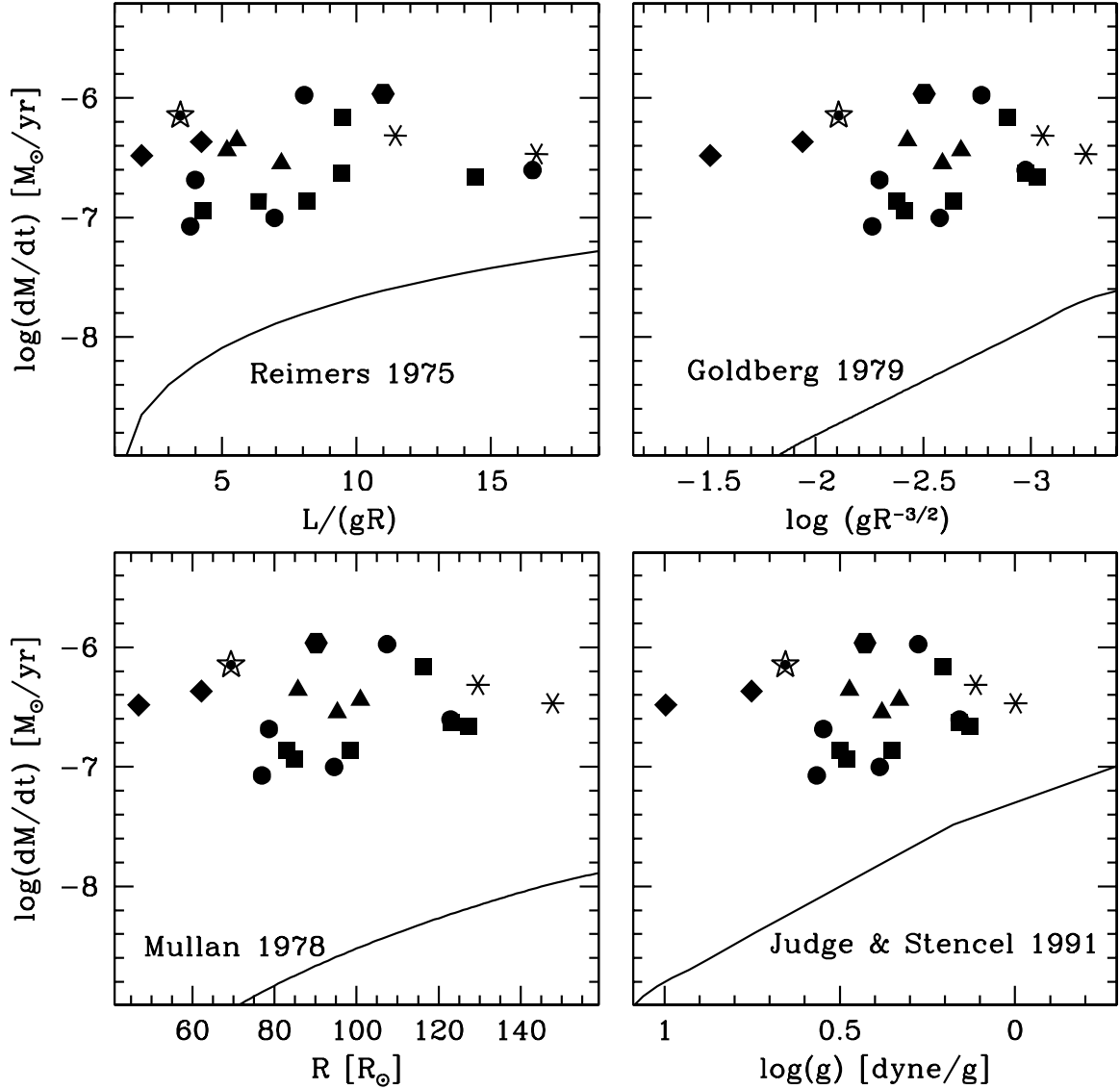}       
\caption{
Mass loss rates for the giant stars with dust excess (cf. Sect.~4), 
as a function of different stellar parameters.  
The position of the V42 long period variable of $\omega$~Cen is also 
marked.  Symbols as in Figs.~3,4.
Different empirical laws by Reimers (1975a,b), Mullan (1978), 
Goldberg (1979), and Judge \& 
Stencel (1991), 
recently revised by Catelan (2000, cf. his Appendix and Figure 10),
calibrated on Population I giants of relatively low luminosity, 
are shown for comparison.
}
\end{figure}  

\end{document}